# Getting to the Bottom of Negative Capacitance FETs

Wei Cao and Kaustav Banerjee
Department of Electrical and Computer Engineering, University of California, Santa Barbara, CA 93106
Contact E-mails: {weicao, kaustav}@ece.ucsb.edu

*Abstract*—In this paper, we take a fresh look at the physics and operation of Negative Capacitance (NC)-FETs, and provide unambiguous feedback to the device designers by examining NC-FETs' design space for sub-60 mV/dec Subthreshold Swing (*SS*). Straightforward design rule is derived, for the first time, based on the capacitor network in NC-FETs. Contrary to many ongoing efforts, it is found that: 1) state-of-the-art MOSFET platforms, such as SOI, FinFET, 2D-FET etc., are not suitable for constructing small-*SS* NC-FETs, unless internal metal gate is introduced; 2) quantum capacitance prevents NC-FETs from achieving hysteresis-free small-*SS*, and low density-of-states (DOS) material can alleviate this issue, to some extent; 3) NC non-linearity can be engineered to reach a tradeoff between sub-60 *SS* and hysteresis; 4) it is more encouraging and practical to use NC to recycle subthreshold voltage loss in short-channel MOSFETs, and overdrive voltage (essentially a type of loss).

## I. INTRODUCTION

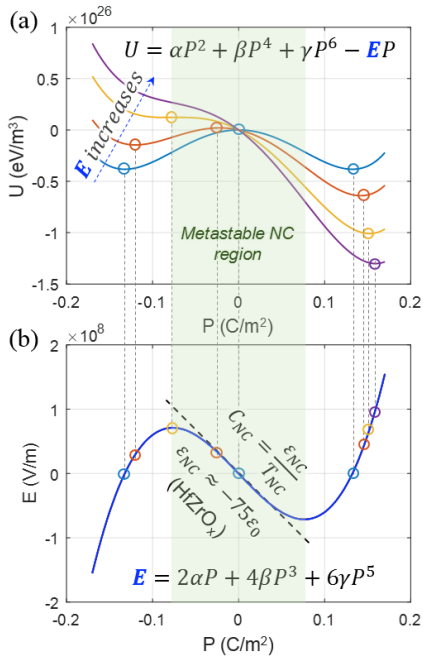

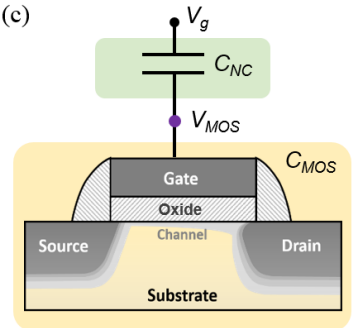

Simply adding a NC layer on top of MOSFET, effective gate voltage can be scaled by a factor of

$$A_v = \frac{\Delta V_{MOS}}{\Delta V_g} = \frac{|C_{NC}|}{|C_{NC}| - C_{MOS}} \geq 1$$

Sounds too easy to be true...

**Fig. 1.** (a) Energy landscape (*U*) of negative capacitance (NC) material in terms of polarization (*P*) and external electric field *E*. (b) Stable (including metastable) *P-E* relation derived from (a). (c) Schematic structure and internal voltage amplification ($A_v$) mechanism of a NC-FET.

The concept of NC is employed to describe the polarization response of "double-well" material systems (typically ferroelectric materials) (**Fig. 1(a)**) in the metastable state, to external electric field, as shown in **Fig. 1(b)** [1]. NC gated FETs are believed to be able to achieve sub-60 *SS*, with minimal penalty in manufacturability, thanks to its simple structure, as shown in **Fig. 1(c)**. Although there have been extensive studies on this topic, many of them, including the original paper in which the idea of NC-FET was proposed [2], have been focused on the interplay between voltage amplification (see the formula in **Fig. 1(c)**) and hysteresis of metal-NC-oxide-metal capacitors, instead of NC-FETs, which have a distinct subthreshold region. Moreover, due to the lack of a straightforward design rule for NC-FETs, experimentalists in this arena have been indiscriminately trying to add NC on various materials/FET structures, ending up with unnecessary failure and/or uninterpretable results. This work is targeted at establishing an unambiguous design framework for NC-FET experimentalists, as well as exploring the design space of NC-FET for hysteresis-free sub-60 *SS*.

## II. ANALYSIS AND DISCUSSION

### A. *Formulating $I_d$-$V_g$ Swing (S)*

The switching of any FET is realized by electrostatic (or capacitive) modulation of the potential of the channel through which current is conducted. For the convenience of analysis, **Fig. 2(a)** depicts all relevant capacitors in a typical NC-FET. $C_Q$, $C_{trap}$, $C_{s/d,geo}$, and $C_{dep}$ are the quantum, trap induced, source/drain geometrical, and depletion capacitances, respectively. These five capacitances can be categorized as gate voltage divider capacitance $C_{div}$, which should be minimized in modern MOSFET design for optimal gate efficiency. The source/drain-to-gate fringing capacitance $C_{fring,s/d-g}$ is irrelevant to channel potential control, and hence to $I_d$-$V_g$ characteristics, as reflected in the small-signal capacitance model in **Fig. 2(a)**. All the other symbols have their nominal connotations. Based on this capacitor network, the swing ($S$) of $I_d$-$V_g$ curve is derived and expressed in the form of decoupled contributions of electrostatics and transport (**Fig. 2(b)**). $S_{TP}$ represents the carrier transport mechanism limited $S$.

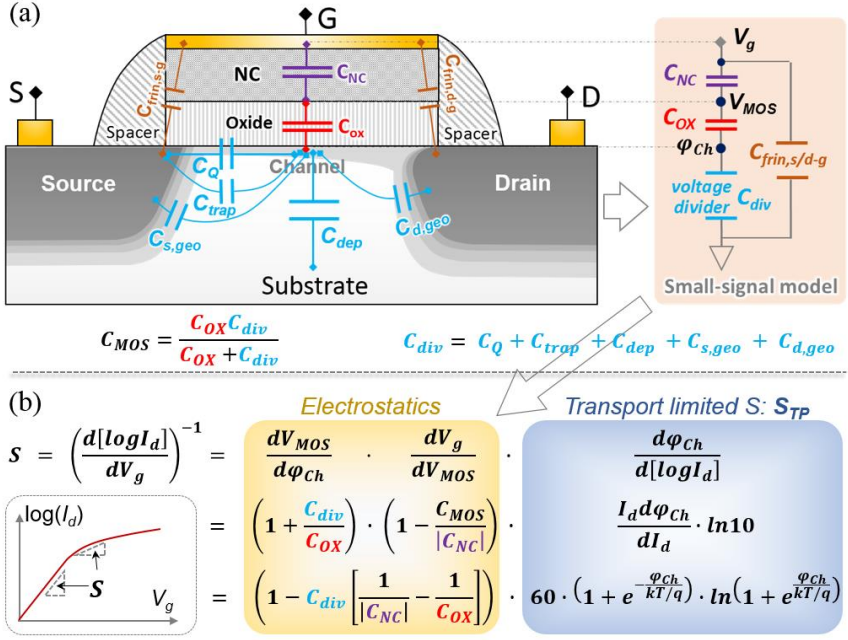

**Fig. 2.** (a) Capacitors in a NC-FET involved in determining the channel potential; $C_Q$, $C_{dep}$, $C_{s/d,geo}$, $C_{trap}$ and $C_{div}$ are the quantum-, depletion-, s/d geometrical, trap induced, and voltage divider capacitors, respectively. (b) Formulating $I_d$-$V_g$ swing ($S$) by decoupling the electrostatics and transport contributions. $S_{TP}$ is the transport limited $S$. Note that $C_{frin,s/d-g}$ does not affect $S$.

### B. Visualizing NC Design Space in Subthreshold Regime

In the subthreshold regime, $S$ is reduced to $SS$. The main difference is the missing mobile charge, and hence $C_Q$. As indicated in the formula in **Fig. 3(a)**, $SS$ is determined by the competition between the body factor ($m$) and NC voltage gain ($A_v$). Only when $|C_{NC}|$ becomes smaller than the gate oxide capacitance $C_{OX}$, $A_v$ can dominate the competition, and deliver sub-60 $SS$, as depicted in **Fig. 3(a)**. In principle, continuously reducing $|C_{NC}|$ can keep reducing $SS$. However, once $SS$ (or $S$) of a NC-FET is designed below zero, the hysteresis will appear in fabricated device, as illustrated in **Fig. 3(b)**. In fact, negative $SS$ (or $S$) in theory and hysteresis in experiments stem from identical physics: polarization switching. Therefore, hysteresis-free subthreshold $I_d$-$V_g$ sets a lower bound on $|C_{NC}|$, which still leaves a wide NC design space, as long as subthreshold divider capacitance $C_{div,<Vth}$ (the slope of the $SS/60$ line) is not designed to be too large.

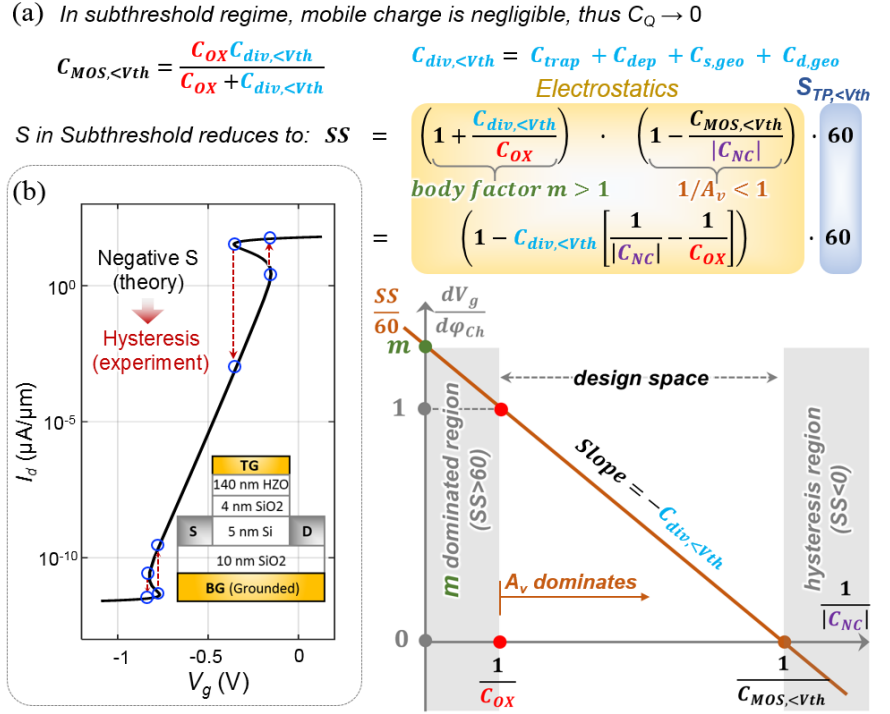

**Fig. 3.** (a) In subthreshold regime, $S_{TP}$ reduces to 60, NC design space for sub-60 subthreshold $S$, i.e. $SS$, and hysteresis-free $I_d$-$V_g$ is $C_{MOS,<Vth}<|C_{NC}|\leq C_{OX}$. The vertical axis $dV_g/d\varphi_{Ch}$ describes gate efficiency. (b) Illustration of the connection between negative $S$ in theory and hysteresis in experimental measurement of $I_d$-$V_g$ curve.

### C. *NC Does Not Help "Good" FETs*

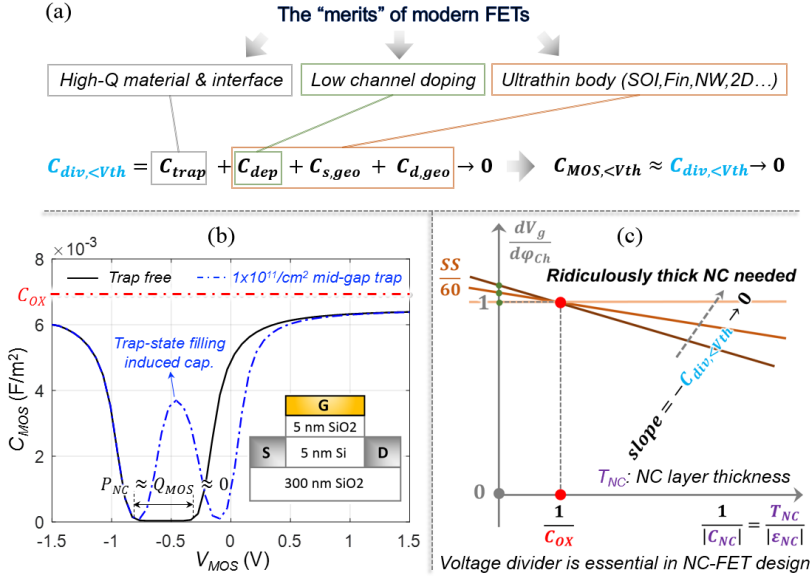

Fig. 4. (a) Modern/"good" FETs have negligible $C_{div,<Vth}$, and hence $C_{MOS,<Vth}$, i.e., no polarization in NC ($P_{NC} \approx Q_{MOS}$), or equivalently $A_v$ benefit, in subthreshold regime. (b) Simulated $C_{MOS}$ in an ultrathin Si SOI MOSFET w/ and w/o traps, quantitatively verifying the analysis in (a). Note that trap states can introduce non-negligible $C_{div,<Vth}$. (c) Illustration of the necessity of voltage divider capacitor.

Modern MOSFETs have evolved into the ultrathin-body (UTB) era, in the form of SOI, FinFET, NWFET, CNT-FET, and 2D-FET etc., accompanied with the requirement of high-quality material/interface and low channel doping, all of which are the constituents of a "good" FET for optimal gate efficiency and current drive (**Fig. 4(a)**). Active experimentalists have often flirted with NC on these "good" FETs, expecting to produce a device with combined inheritance [3][4]. However, the merits of these modern FETs make $C_{div,<Vth}$ ($=C_{trap}+C_{dep}+C_{s,geo}+C_{d,geo}$), and hence $C_{MOS,<Vth}$ negligibly small (**Fig. 4(a)**), which is quantitatively verified in the simulated $C_{MOS}$ of a Si SOI MOSFET (**Fig. 4(b)**). As a result, it is hard to achieve sub-60 $SS$, unless unfeasibly thick NC layer is used, as illustrated in **Fig. 4(c)**. From NC operation point of view, subthreshold charge density $Q_{MOS,<Vth}$ ($=C_{MOS,<Vth} \cdot V_g$) in these "good" FETs induces (through electric displacement field $D$ ($=C_{MOS,<Vth} \cdot V_g$), which is continuous across oxide/NC interface) negligible polarization in NC, and thus miniscule voltage amplification $A_v$. Therefore, contrary to MOSFET design, a voltage divider (i.e., finite $C_{div,<Vth}$) should be intentionally introduced into NC-FETs to exploit $A_v$ benefits.

### D. *Quantum Capacitance "Kills" Sub-60 SS*

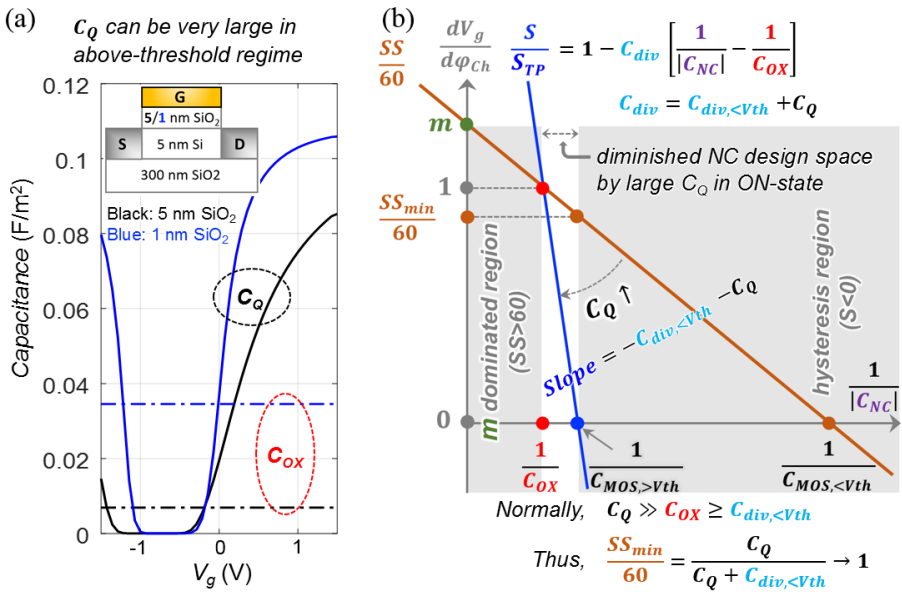

Fig. 5. (a) Simulated $C_Q$ in an ultrathin Si SOI MOSFET. Even in such a thin film, $C_Q$ is still much larger w.r.t. the 1 nm SiO$_2$ gate oxide capacitance $C_{OX}$. (b) Hysteresis should be eliminated not only in subthreshold regime as the design in **Fig. 3(b)**, but also in above-threshold regime. The rapidly increasing $C_Q$ with gate bias turns the $S/S_{TP}$ curve clockwise, narrowing the NC design space. Eventually, the ultra-large $C_Q$ in the above-threshold regime can diminish the NC design space, and constrain the obtainable minimum $SS$ ($SS_{min}$) close to 60 mV/dec. Note $C_{div}=C_{div,<Vth}+C_Q$, which is essentially the slope of $S/S_{TP}$ curve.

In FETs, quantum capacitance $C_Q$ describes the response of the mobile charges to channel potential modulation. In contrast to $C_{div,<Vth}$ (and hence $C_{MOS,<Vth}$), as well as $C_{OX}$, $C_{div}$ (and $C_{MOS}$) in near- and above-threshold regime are greatly contributed by mobile charge carriers, i.e., $C_Q$. Even in the ultrathin channel of a UTB SOI MOSFETs, $C_Q$ is still much larger than $C_{OX}$ (**Fig. 5(a)**). It increases the slope of $S/S_{TP}$ line, i.e., rotates the $S/S_{TP}$ line clockwise, w.r.t. the $SS/60$ line, as illustrated in **Fig. 5(b)**, and significantly raises the lower bound of $|C_{NC}|$ (below which hysteresis appears), eventually diminishing the NC design space (in between $C_{OX}C_Q/(C_{OX}+C_Q)$ and $C_{OX}$). The minimum achievable $SS$ ($SS_{min}$) is raised to 60 $C_Q/(C_Q+C_{div,<Vth})$, which approaches 60, in normal MOSFETs.

### E. *Designing NC-FET in the Quantum Capacitance Limit*

Based on above analysis, lowering $C_Q$, desirably into the quantum capacitance limit ($C_Q<C_{OX}$), seems an effective direction, in enlarging NC design space and reaching small $SS_{min}$. To achieve low $C_Q$, low-DOS material systems could help. $SS_{min}$ and NC design space are evaluated for Si, Ga$_{0.47}$In$_{0.53}$As, and 2D MoS$_2$, in **Fig. 6(a)** and **(b)**, respectively, versus $C_{div,<Vth}$ and $C_{OX}$, which are normalized to equivalent oxide thickness (EOT) for the convenience of comparison. For bulk materials, $C_{div,<Vth}$ can be realized in the form of $C_{dep}$, by engineering the channel doping level and profile. For 2D materials which lack effective and controllable doping techniques, a bottom gate can be introduced to serve as $C_{div,<Vth}$ (=$C_{BOX}$). As shown in **Fig. 6(a)(b)**, the low-DOS Ga$_{0.47}$In$_{0.53}$As does provide much smaller $SS_{min}$, and larger NC (HfZrOx is used in the calculation) design space (difference in NC thickness at $SS_{min}$ and $SS$ of 60), w.r.t. Si and MoS$_2$. Further reduction of the DOS of GaInAs system by increasing In content is feasible, but will result in lowered bandgap that degrades ON-OFF current ratio. The emerging 2D MoS$_2$, although promising for low-power MOSFETs due to its thinness [5], has relatively large electron/hole effective masses, which makes it a large-DOS system for FETs. In contrast, Germanane (a single-layer 2D semiconductor) possesses small effective mass and decent bandgap [6], thus can benefit NC-FET design. **Fig. 6(c)** shows simulated $I_d$-$V_g$ curves of Si, Ga$_{0.47}$In$_{0.53}$As, and Germanane NC-FETs. With exactly the same structure/size (inset), hysteresis appears in the Si NC-FET but absent in the Ga$_{0.47}$In$_{0.53}$As device, because of the low-DOS benefit. Optimized Germanane NC-FET can achieve hysteresis-free 39 mV/dec $SS$.

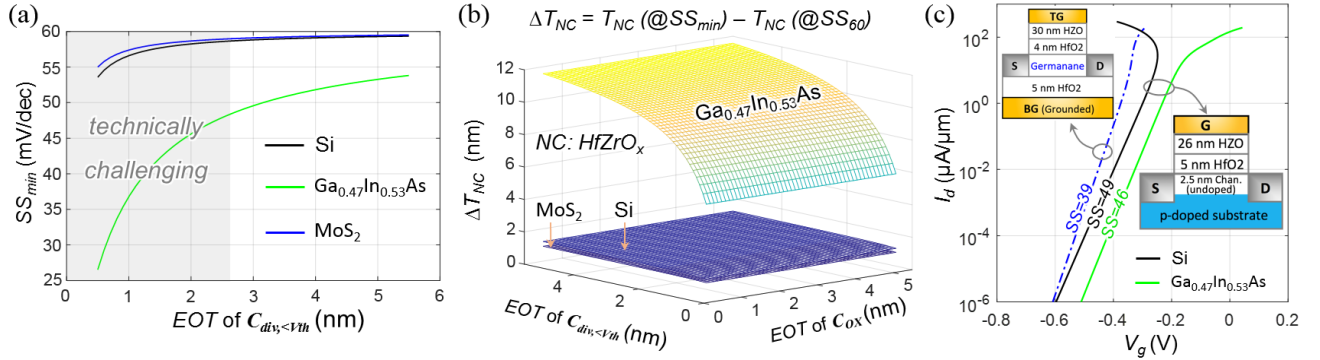

**Fig. 6.** Calculated **(a)** $SS_{min}$ and **(b)** NC thickness design space ($\Delta T_{NC}$) for Si, Ga$_{0.47}$In$_{0.53}$As, and MoS$_2$. The low-*DOS* Ga$_{0.47}$In$_{0.53}$As shows great advantages. $C_{div,<Vth}$ and $C_{OX}$ are normalized to equivalent oxide thickness (EOT). **(c)** With the same device structure/size, Ga$_{0.47}$In$_{0.53}$As device can achieve smaller $SS$, w.r.t. Si device, without hysteresis. $SS$ of Germanane device can reach 39 mV/dec.

### F. *The Role of NC Non-Linearity*

Constant NC dielectric constant $|\varepsilon_{NC}|$ or $|C_{NC}|$, i.e., linear $P$-$E$ relation, is valid only within limited range (**Fig. 1(b)**), it increases with $Q_{MOS}$ or gate bias, as indicated by the derived formula in **Fig. 7(a)**. This non-linearity introduces a damping term to the slopes of both $SS/60$ and $S/S_{TP}$ lines versus NC layer thickness $T_{NC}$ (**Fig. 7(a)**), which can be depicted as a counterclockwise rotation of both $S/S_{TP}$ and $SS/60$ lines, leading to enlarged NC design space, but increased $SS$, as illustrated in **Fig. 7(b)**. **Fig. 7(c)** and **(d)** show simulated $I_d$-$V_g$ and $S$-$I_d$ characteristics of a Si SOI NC-FET (inset), respectively, with different non-linearity term $\beta$. With increased $\beta$, the hysteresis is effectively eliminated, at the expense of increased $SS$, which is consistent with the uncovered role of NC non-linearity in **Fig. 7(b)**. A balance between sub-60 $SS$ and NC stability can be achieved in principle by engineering this non-linearity.

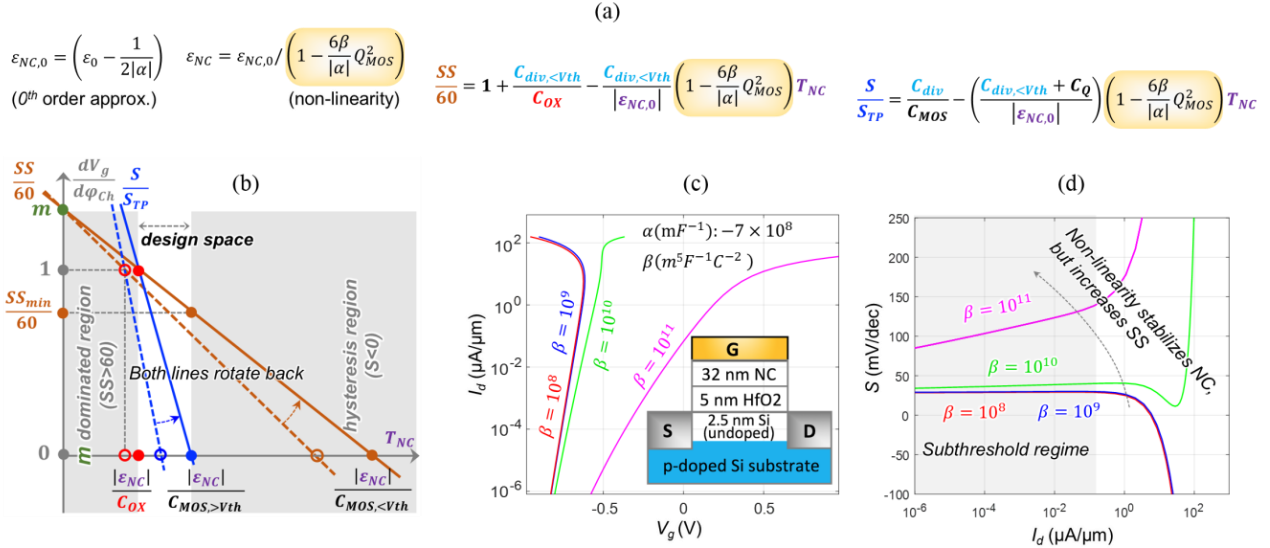

**Fig. 7.** (a) The non-linearity of NC dynamically increases $|\varepsilon_{NC}|$ (or $|C_{NC}|$) with $Q_{MOS}$, or gate bias. (b) It rotates back both $S/S_{TP}$ and $SS/60$ lines, which enlarges the NC design space constrained by $C_Q$, but increases $SS$. (c),(d) Verification of the uncovered non-linearity physics by varying the strength of NC non-linearity, $\beta$, in simulation.

### G. IMG: Borrow Parasitic Charge for Polarization in NC

An internal metal gate (IMG) inserted between the NC layer and oxide have been found able to achieve small $SS$ in short-channel FinFETs, without introducing hysteresis [7]. This section clarifies the underlying physics using the same framework developed above. **Fig. 8(a)** shows the capacitor network in a NC-FET with IMG. It can be found that two additional fringing (overlapping included) capacitors, $C_{frin,s/d\text{-}IMG}$, appear between source/drain and the floating IMG. $S$ formula is revisited for this structure, as shown in **Fig. 8(b)**. For those "good" FETs, such as FinFETs, $C_{div,<Vth}$ is negligible, and $S$ and $SS$ formula can be reduced to very simple forms (**Fig. 8(b)**), and are plotted in **Fig. 8(c)**. Interestingly, the NC design space and $SS_{min}$ become irrelevant w.r.t. $C_Q$, and only dependent on $C_{OX}$ and $C_{frin,s/d\text{-}IMG}$. This is because $C_{frin,s/d\text{-}IMG}$ provide parasitic charge in both subthreshold and above-threshold regime to induce polarization in NC, which unlocks the dependence of polarization on mobile charge (or $C_Q$) in "good" FETs. As indicated by the $SS_{min}$ formula inside **Fig. 8(c)**, $SS_{min}$ can be effectively reduced by increasing the ratio of $C_{frin,s/d\text{-}IMG}$ to $C_{OX}$, which perfectly explains the previous report [7] that short-channel NC FinFETs with IMG can achieve small *SS*. For long-channel NC-FETs, large overlap between source/drain and IMG, w.r.t. channel length, can be designed to reduce *SS*, which is confirmed by simulation of a Si SOI NC-FET, as shown in **Fig. 8(d)**. However, it is worth noting that introducing IMG into NC-FET gate stack may induce floating gate memory effect.

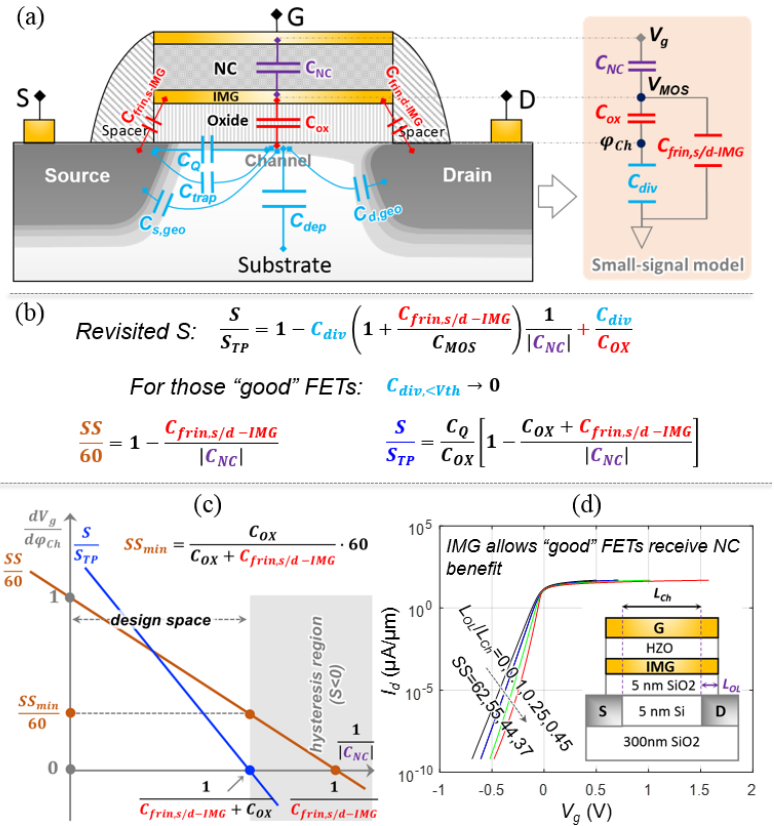

**Fig. 8.** (a) Capacitors in a NC-FET with internal metal gate (IMG). (b) Revisited $S$ and $SS$ formula. (c) IMG can borrow charge from fringing/overlap capacitance in subthreshold regime, thereby unlocking polarization from $C_Q$, and enlarging the NC design space for small $SS_{min}$. (d) Simulation results based on a Si SOI MOSFET verify the uncovered the IMG mechanism. The larger the $L_{OL}/L_{Ch}$, more the NC benefits.

### H. *A More Practical Role of NC for FETs - Voltage Loss Recycler*

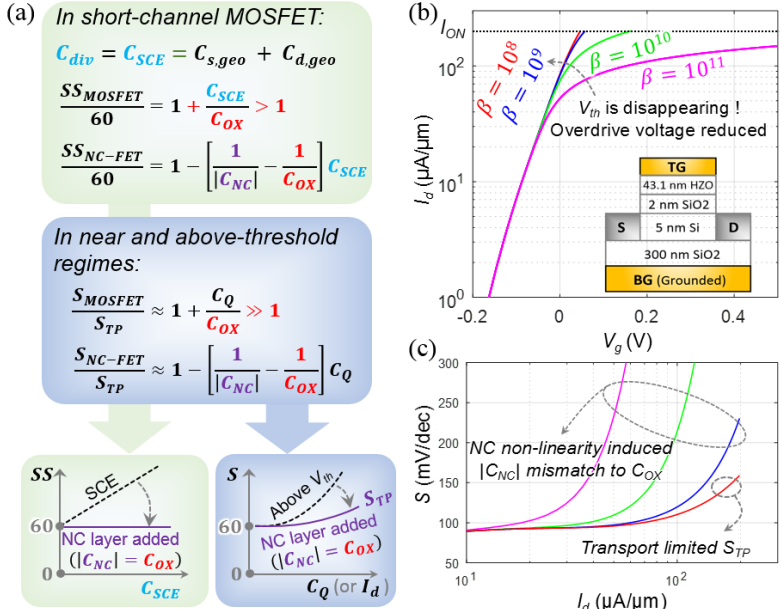

**Fig. 9.** (a) With a simple $|C_{NC}|=C_{OX}$ matching, *SS* can be restored to 60 mV/dec in short-channel MOSFETs, and *S* above threshold can be reduced to the transport limit ($S_{TP}$). Thereby, subthreshold voltage loss and overdrive voltage loss are recycled. (b)(c) Verification of the idea with simulation.

Although sub-60 mV/dec *SS*, as discussed above, can be achieved in NC-FETs in principle, non-trivial engineering efforts are needed. In fact, there is a more practical and encouraging role of NC for FETs. **Fig. 9(a)** provides a *SS* formula for short-channel MOSFETs and NC-FETs, and a *S* formula in the above-threshold regimes for MOSFETs and NC-FETs of any channel length. In the short-channel MOSFETs, short-channel effects (SCE) causes a portion of subthreshold gate voltage loss on the gate oxide, which is reflected by the non-unity gate efficiency, $1+C_{SCE}/C_{OX}>1$. In above-threshold regime, large $C_Q$, or charge screening effect, forces $V_g-V_{th}$, which is usually called overdrive voltage, to drop on the gate oxide, instead of in the channel, leading to the poor gate efficiency, $1+C_Q/C_{OX}\gg1$. In this sense, overdrive voltage is a type of voltage loss. Interestingly, by adding a NC layer on MOSFETs, and simply matching $|C_{NC}|$ to the constant $C_{OX}$, both $1+C_{SCE}/C_{OX}$ and $1+C_Q/C_{OX}$ are absorbed, no matter how large $C_{SCE}$ or $C_Q$ is, as illustrated in the schematic in **Fig. 9(a)**. In other words, *S* can be restored to 60 in subthreshold, and to transport mechanism limited $S_{TP}$ in the above-threshold regimes. The significance of the latter is that it provides an alternative solution to reduce the supply voltage and hence switching energy, i.e., by saving overdrive voltage, which also occupies a big portion of the entire supply voltage, instead of struggling to save subthreshold voltage. This idea is supported by the simulation results in **Fig. 9(b)** and **(c)**. Note that small NC non-linearity is desired to reduce *S* in above-threshold regime. Finally, it is instructive to note that polarization rate in NC [8] should be improved to GHz range, for high-performance and low-power logic application.

### III.  SUMMARY

A summary of this work, as well as relevant suggestions to NC-FET experimentalists, are provided as follows:

1) NC does not help state-of-the-art MOSFET platforms (such as SOI, FinFET, NWFET, CNT-FET, 2D-FET…) achieve sub-60 mV/dec *SS*, unless internal metal gate (IMG) and overlap capacitance (between source/drain and IMG) are introduced. However, IMG may result in floating gate memory effect and lowered thermal budget issue.

2) Appropriate voltage divider capacitor (generally minimized in conventional MOSFET design) should be intentionally introduced to exploit NC benefit in the subthreshold regime.

3) Smaller-density-of-states (DOS) materials provide larger NC design space as well as smaller $SS_{min}$. Maximum benefit can be derived from NC FETs working in the quantum capacitance limit.

4) The non-linearity of NC helps reduce hysteresis, but leads to increased *SS*. A balance between small *SS* and hysteresis-free I-V can be reached in principle by engineering this non-linearity.

5) The varying quantum capacitance makes $|C_{NC}|$-to-$C_{MOS}$ matching very difficult. From a more practical point of view, it is encouraging to simply match $|C_{NC}|$ to the constant gate oxide capacitance, which can effectively recycle the subthreshold voltage loss of short-channel MOSFETs, and reduce the overdrive voltage (charge screening effect forces overdrive voltage to drop primarily across the gate oxide, which is essentially a loss).